\documentclass[aps,preprint,showpacs,amsfonts]{revtex4-1}
\usepackage[utf8x]{inputenc}
\usepackage[T1]{fontenc}
\usepackage{textcomp}
\usepackage{amsmath}
\usepackage{latexsym}
\usepackage{float}
\usepackage{amsfonts}
\usepackage{accents}
\usepackage{amssymb}
\usepackage{graphicx}
\usepackage{hyperref}
\usepackage{subfigure}
\usepackage{bbold}
\usepackage{xcolor}
\usepackage{url}
\usepackage{booktabs,tabularx,dcolumn}

\begin{document}
\title{Role of temperature and alignment activity on kinetics of coil-globule transition of a flexible polymer}

\author{Subhajit Paul} \email[]{subhajit.paul@itp.uni-leipzig.de}
\author {Suman Majumder}\email[]{suman.majumder@itp.uni-leipzig.de}
\author{Wolfhard Janke}\email[]{wolfhard.janke@itp.uni-leipzig.de}
\affiliation{Institut  f\"{u}r Theoretische Physik, Universit\"{a}t Leipzig, IPF 231101, 04081 Leipzig, Germany}
\date{\today}



\begin{abstract}
We study the nonequilibrium kinetics during the coil-globule transition of a flexible polymer chain with active beads after a quench from good to poor solvent condition using molecular dynamics simulation. Activity for each bead is introduced via the well-known Vicsek-like alignment rule due to which the velocity of a bead tries to align towards the average direction of its neighbors. We investigate the role of quenching temperature with varying activity during collapse of this polymer. We find that although for lower activities the kinetics remains qualitatively similar for different temperatures, for higher activity noticeable differences can be identified.
\end{abstract}
\maketitle
\section{Introduction}
\par 
The typical conformation of a  polymer undergoes changes when quenched from a good to a poor solvent condition \cite{doi}. In a good solvent or at high temperature the equilibrium conformation of the polymer is a coil. On the other hand, in a poor solvent or at low temperature the equilibrium conformation becomes a globule. The equilibrium aspects of this transition have been studied since many years and are quite well understood \cite{doi}. Despite efforts using computer simulations of coarse-grained as well as all-atom models, a concrete theoretical framework for the nonequilibrium properties is still missing.  Among different phenomenological descriptions regarding the pathways of globule formation, the `pearl-necklace' picture proposed by Halperin and Goldbart \cite{halperin} is quite well accepted and observed in various polymeric systems. In recent years, the kinetics and associated scaling laws have been explored by using the analogy with usual coarsening in a spin or particle system \cite{majumder1,majumder2,henrik}.

\par 
Understanding of relaxation across the collapse transition for a polymer has relevance in many systems, e.g., protein folding, chromatin dynamics and conformational changes of other bio-polymers, to name a few. Such polymer chains can also be `active' in nature. 
Their `activity' can be modeled using frameworks borrowed from `active' particle systems \cite{elgeti}. Active particles have the property of being motile either by using their own internal
energy or by taking energy from the environment.  Similarly, for an active polymer, the monomers can be active by themselves or can be activated 
by external forces. Like in a particle system, for a polymer
also one should expect significant differences in its dynamics due to activity compared to that of a passive polymer. 
Whereas notable progress has been made for a passive polymer \cite{majumder1,majumder2,henrik},
studies regarding the kinetics of an `active' polymer are rather recent \cite{winkler,bianco,paul1,paul2}.

\par 
In this paper, we consider a flexible homopolymer chain with each monomer as an active element. The activity is implemented via the well-known Vicsek-like alignment rule \cite{paul1,paul2,vicsek,das1,paul3}. In this context a polymer can be visualized as a system in which the beads are moving with some constraints. Due to this activity, the velocity of each bead gets modified towards the average direction of its neighbors. In our earlier works \cite{paul1,paul2}, we have investigated the effects of such an activity on the pathway across the coil-globule transition and compared them with those for the passive case for a particular  quench temperature. Here we consider a lower quench temperature and investigate whether similar features can be observed for the conformations and kinetics.

\section{Model and methods}
We consider a bead-spring polymer chain with $N$ linearly connected beads. The bonded interaction between two successive beads is modeled via a finitely extensible non-linear elastic (FENE) potential \cite{majumder1} in the form
	$V_{\rm{FENE}}(r) = - 0.5 KR^2 {\rm{ln}} \big[ 1- \big((r-r_0)/{R}\big)^2\big]\,$ with $K=40$, $r_0=0.7$ and $R=0.3$. The non-bonded interaction is modeled as $V_{\rm{nb}}(r_{ij})=V_{\rm{LJ}}(r)-V_{\rm{LJ}}(r_c) -(r-r_c)\frac{dV_{\rm{LJ}}}{dr}$, where
	$V_{\rm{LJ}}(r) = 4\epsilon \big[\big({\sigma}/{r}\big)^{12}- \big({\sigma}/{r}\big)^6\big]\,$ is the standard Lennard-Jones (LJ) potential. Here  $\sigma=r_0/2^{1/6}$ is the diameter of the beads, $\epsilon=1$ the interaction strength and $r_c=2.5\sigma$ is the cut-off distance.
\par
Dynamics of this polymer has been studied using molecular dynamics (MD) simulation with velocity-Verlet integration scheme using the Langevin thermostat. Thus for each bead our working equation is \cite{paul1,paul2}
\begin{equation}\label{lang}
	m \frac{d^2{\vec{r}}_i}{dt^2} = - \vec{\nabla} U_i - \gamma \frac{{d\vec{r}}_i}{dt} + \sqrt{2 \gamma k_B T } \vec{\Lambda}_i(t)\,,
\end{equation}
where $m$ is the mass,  $\gamma$ the drag coefficient, $k_B$ the Boltzmann constant, $T$ the quench temperature measured in units of $\epsilon/k_B$, and $U_i$ is the interaction potential which contains both $V_{\rm{FENE}}$ and $V_{\rm{nb}}$. For convenience we set $m$, $\gamma$ and $k_B$ to unity.  $\Lambda_i(t)$ represents Gaussian white noise with zero mean and unit variance, and Delta-correlations over space and time. We have used the integration time step $\delta t=0.0005$ in units of the timescale $\tau_0=\sqrt{m\sigma^2/\epsilon}$.
\par 
At each MD step the activity is introduced in the following way. The direction of the velocity of the $i$-th bead obtained from Eq.~(\ref{lang}) is modified by the active force \cite{paul1,paul2,paul3}
\begin{equation}
	\vec{f}_i =f_A\hat{n}_i\,;~~~~{\rm{with}} ~~\hat{n}_i = \frac{\big(\sum_j \vec{v}_j\big)_{r_c}}{\big|\big(\sum_j \vec{v}_j\big)_{r_c}\big|}\,,
\end{equation}
 where $f_A$ is the strength of the activity and $\hat{n}_i$ represents the average direction of the neighboring beads within a sphere of radius $r_c$. More technical details regarding the implementation of the active force are discussed in Refs.~\cite{paul1,paul2}. 
\par 
In the rest of the paper, the activity strength will be expressed via the ratio of active (``ballistic'') and thermal energy, i.e., the P\'eclet number \cite{winkler}
\begin{equation}
	Pe=\frac{f_A\sigma}{k_BT}.
\end{equation}
As we will focus on comparing the kinetics of the polymer collapse with activity for different temperatures, it is more convenient to use the dimensionless parameter $Pe$ for which we consider the values $0,~0.62,~1.25$ and $5.0$. $Pe=0$ corresponds to the passive polymer case. The temperatures are chosen as $T=0.25$ and $0.5$ which are both well below the coil-globule transition temperature for the passive polymer of length $N=512$ \cite{majumder2}. In the figures of the next section, $\langle \dots \rangle$ represents an average over different initial configurations and thermal noise, using $200$ independent  realizations for $T=0.25$ and $500$ for $T=0.5$.
\section{Results}
\par
\begin{figure}[b!]
	\centering
	\includegraphics*[width=15cm, height=12.0cm]{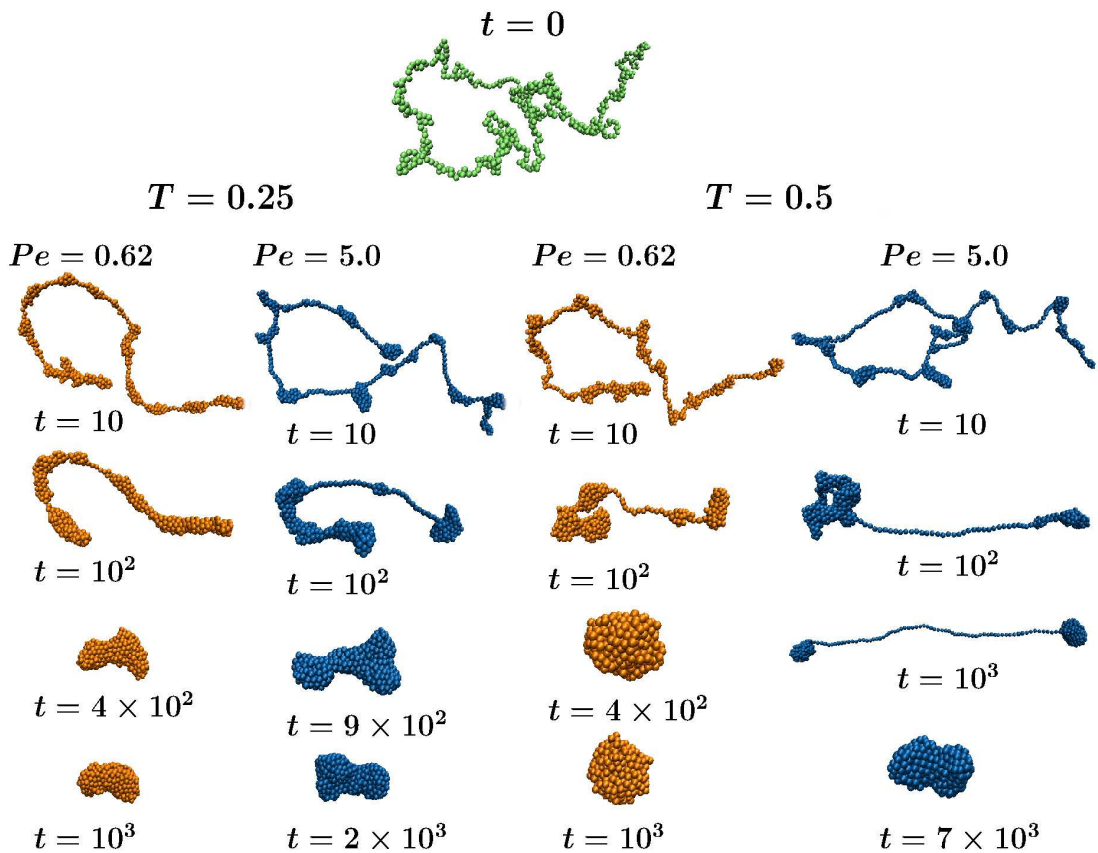}
	\caption{\label{snap} Typical snapshots representing the time evolution during collapse of an active polymer of length $N=512$ at two different temperatures for two different values of the P\'eclet number $Pe$.}
\end{figure}

 
 We start our discussion by presenting in Fig.~\ref{snap} typical representative conformations of the polymer chain during its evolution with time for the two temperatures at  low  and high activity, i.e.,  $Pe=0.62$ and $5.0$, respectively. Comparative time evolution snapshots for different activities as well as for the passive case are discussed in Ref.~\cite{paul2}. Although in all four cases the final conformations are globules, one notices differences in the pathways. For $Pe=0.62$ the conformations follow the three different stages of the `pearl-necklace' picture \cite{halperin}.
For $Pe=5.0$ and $T=0.25$ the time evolution does not look very much different, although it takes longer time for the globule to form. For $T=0.5$, however, already the early time conformations look somewhat different. It appears that the polymer becomes more elongated than its starting conformation and then starts collapsing. Later the two-cluster conformations, i.e., the dumbbell state, persists quite long and it takes very long to reach the collapsed state. Even though the value of $Pe$ is the same, the conformations as well as the time required for globule formation at these two different temperatures are clearly different.
\par 
 Before going into the quantification of conformational properties or the cluster growth, we first look at the effect of increasing activity on the alignment of the velocities of the beads. Thus we define a velocity order parameter as
\begin{equation}\label{opv}
 v_a = {\mid\sum\limits_{i=1}^{N} \vec{v}_i\mid}\Big/{\sum\limits_{i=1}^{N} \mid\vec{v}_i\mid}\,,
\end{equation}
where $\vec{v}_i$ is the velocity of the $i$-th bead. In Figs.~2(a) and (b) we show plots of $\langle v_a \rangle$ vs time for different values of $Pe$ for the two temperatures. For both of them, in the passive case, $\langle v_a \rangle$ always remains close to $0$. Then for $Pe > 0$, $\langle v_a \rangle$ saturates at non-zero values which increase with increasing $Pe$. But for the same $Pe$ the saturation values differ for the two temperatures. In fact, with $Pe=5.0$ one finds a notable difference. For $T=0.25$, $\langle v_a \rangle$ increases smoothly and reaches its saturation at $\approx 0.8$, whereas for $T=0.5$ the saturation occurs in two steps. The slower growth after the first plateau is due to the longer persistence of the dumbbell conformation for which the two clusters move slowly towards each other before they finally merge and $\langle v_a \rangle$ reaches up to $0.9$ \cite{paul2}. Even though the $Pe$ values for the two temperatures are the same, this does not lead to the same degree of alignment of the beads.
\par 
The conformational changes during the collapse can be quantified via the squared radius of gyration $R_g^2$ of the polymer defined as
\begin{equation}\label{rgsq_def}
R_g^2 = \frac{1}{N}\sum_{i=1}^{N} (\vec{r}_{\rm{cm}} - \vec{r}_i)^2\,,
\end{equation}
where $\vec{r}_{\rm{cm}}=\frac{1}{N}\sum_{i=1}^{N} \vec{r}_i$ defines the center-of-mass of the polymer.
\begin{figure}[t!]
	\centering
	\includegraphics*[width=13.0cm, height=6.0cm]{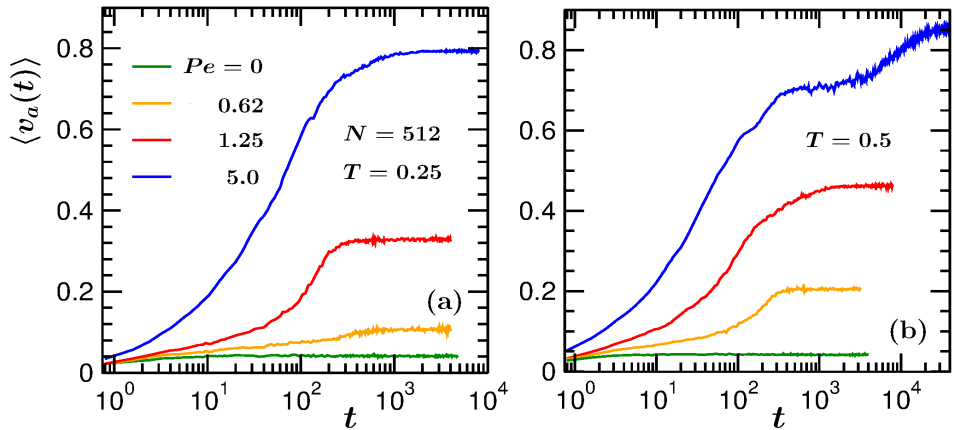}
	\caption{\label{velo_opm}  Semi-log plots of the average order parameter $\langle v_a(t) \rangle$ versus $t$ for two different quench temperatures $T$ of a polymer of length $N=512$. For both temperatures, data are shown for different values of the P\'eclet number $Pe$.}
\end{figure}
To show a comparative picture, in Figs.~\ref{rg_t}(a) and (b) we plot $\langle R_g^2 \rangle$ versus $t$ for $T=0.25$ and $T=0.5$, respectively. For $T=0.5$ with low values of $Pe$ the decay is faster than that for the passive case. For $Pe=5.0$ we see that $\langle R_g^2 \rangle$ initially increases from its starting point and then follows a much slower decay. 
For $T = 0.25$ and lower $Pe$ we observe a similar trend of the decay of $\langle R_g^2 \rangle$, but here it even becomes faster with increasing $Pe$. For larger activity with $Pe = 5.0$ we do not observe the initial increase of $\langle R_g^2 \rangle$ but as for lower $Pe$ a faster decay than in the passive case. Only asymptotically for large $t$, similar to $T=0.5$, the decay appears to become slower.
For a better visualization we plot the data for $Pe=5.0$ for both $T$ values over a much longer time on a semi-log scale in Fig.~\ref{rg_t}(c). This clearly shows  that even though  $Pe$ is kept at the same value, the conformational changes during the kinetics of globule formation are not similar. This has also been observed from the corresponding snapshots in Fig.~\ref{snap}.  From this one can conclude that the P\'eclet number is not the only determining parameter for the kinetics.
\begin{figure}[t!]
	\centering
	\includegraphics*[width=16.50cm, height=5.8cm]{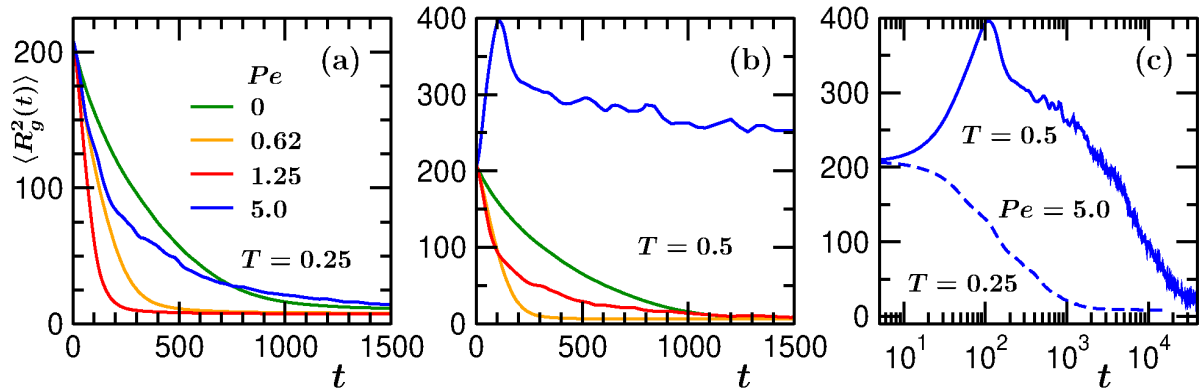}
	\caption{\label{rg_t} Plots of the average squared radius of gyration $\langle R_g^2(t) \rangle$ versus time $t$ for passive as well as active polymers with $N=512$ for (a) $T=0.25$ and (b) $T=0.5$. In both plots the values of $Pe$ are the same. (c) Semi-log plot of $\langle R_g^2(t) \rangle$ versus $t$ for the higher activity, i.e., with $Pe=5.0$ for both the temperatures.}
\end{figure}
\par
\begin{figure}[t!]
	\centering
	\includegraphics*[width=13.0cm, height=6.50cm]{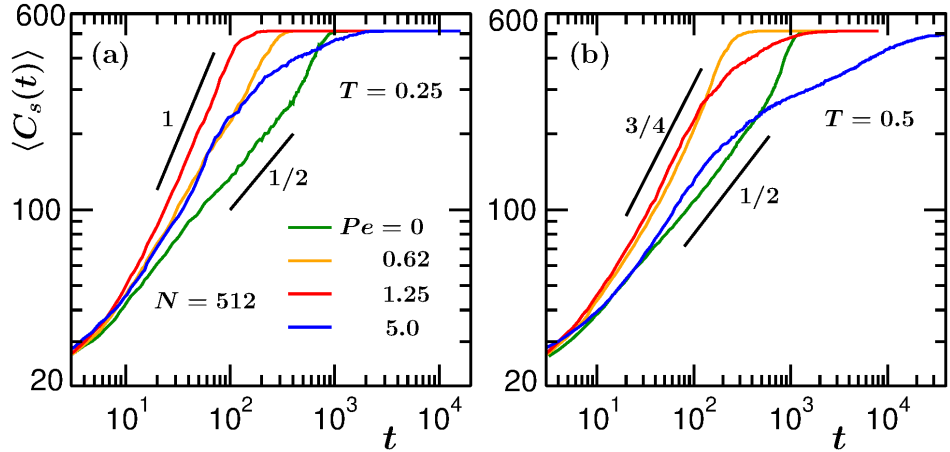}
	\caption{\label{cluster}  Log-log plots of $\langle C_s(t) \rangle$ versus $t$ for a polymer with $N=512$ governed by different values of $Pe$ for (a) $T=0.25$ and (b)  $T=0.5$. In both plots the exponents of the corresponding power laws are mentioned close to the curves.}
\end{figure}

 
 Finally we investigated whether there exist any differences in the cluster growth at these two temperatures. For this we have identified the number of clusters $n_c(t)$ at time $t$ along the chain and calculated their average size as $C_s(t) = \big(\sum_{k=1}^{n_c(t)} m_k\big)/n_c(t)$ where $m_k$ counts the number of monomers in the $k$-th cluster. More technical details can be found in Refs.~\cite{majumder1,majumder2,paul2}. In Figs.~\ref{cluster}(a) and (b) we plot $\langle C_s(t) \rangle$ versus $t$ for different values of $Pe$ for both temperatures. In general, $\langle C_s(t) \rangle$ follows a power-law behavior with time 
\begin{equation}
\langle C_s(t) \rangle \sim t^{\alpha_c}\,,
\end{equation}
where $\alpha_c$ is the growth exponent. We see that for both temperatures the qualitative behavior of $\langle C_s(t)\rangle$ with increasing $Pe$ remains quite similar. For the passive case, in the scaling regime data look consistent with $\alpha_c \approx 1/2$ for both values of $T$. In both cases, with lower activities the growth is faster and the globular state is reached earlier than in the passive case. But the exponent for $T=0.5$ appears to be lower than for $T=0.25$. As guide to the eyes, we plot power-law lines with exponents $1$ and $3/4$ for $T=0.25$ and $0.5$, respectively. For $Pe=5.0$ we see an opposite trend as the growth becomes slower and it takes a  much longer time to reach the final globular state. In fact, similar to the lower activities, for this higher $Pe$ also, it appears that the growth is slower for $T=0.5$ compared to $T=0.25$.  
\par 

\section{Conclusion}
In this paper we have presented results for the kinetics during the coil-globule transition of a flexible polymer following a  quench from a good to a poor solvent condition. The beads for the polymer are considered to be `active' which is implemented in a Vicsek-like alignment manner. For a comparison among two different quench temperatures we keep the values of the P\'eclet number $Pe$ the  same. For the passive case as well as for low $Pe$ values, the conformations and kinetics look more or less similar. Noticeable differences in the structural behavior and the kinetics appear for the higher activity. Our observations indicate that $Pe$ cannot be the only dimensionless parameter determining the effect of activity. Rather, one may need to consider also another energy scale, namely  the interaction energy $\epsilon$. Investigation of the interplay of different energy scales in detail is beyond the scope of this paper and will be presented elsewhere in the future.

\section*{Acknowledgments}
This project was funded by the Deutsche Forschungsgemeinschaft (DFG, German Research Foundation) 
under Grant No.\ 189\,853\,844--SFB/TRR 102 (Project B04). It was further supported by the Deutsch-Franz\"osische Hochschule (DFH-UFA) through the Doctoral College ``$\mathbb{L}^4$'' under Grant No.\ CDFA-02-07, the Leipzig Graduate School of Natural Sciences ``BuildMoNa'', 
and the EU COST programme EUTOPIA under Grant No.\ CA17139.

\end{document}